# Deformation Effect on Graphene Quantum Dot/Graphane and Silicene Quantum Dot/Silicane array

*Bi-Ru Wu*

Department of Natural Science, Center for General Education, Chang Gung University, No.259, Wenhua 1st Rd., Guishan Dist., Taoyuan City 33302, Taiwan.

ABSTRACT:

This article presents a design for the two-dimensional heterostructure (2DH) systems of graphane quantum dot array in graphane (GQD/Graphane), and silicene quantum dot array in silicane (SiQD/Silicane). A first-principles method was used to evaluate the deformation effect for magnetism as well as the electronic properties for the 2DH systems. The energy levels of quantum dot (QD) array and the band structure of its hydrogenated counterpart are coupling for both 2DH systems of C and Si. The hydrogenated part shares part of strain on QD array, however, the strain sharing effect is stronger in SiQD/silicane than in GQD/graphane. The strain sharing enhances the band coupling of the QD and its hydrogen counterpart in the low energy region. The band coupling alters the electronic properties of the 2DH systems and change the magnetic properties of triangular and parallelogram of SiQD/Silicane array under compressive strain larger than 5%. Strain modulates the band gap of the 2DH system.  For SiQD/Silicane systems, the homogeneous strain not only induces the phase transition of from semiconductor to metal, but also remove the magnetism of triangular and parallelogram SiQD array. The 2DH system can be used in the design of nanoelectronic devices and binary logic based on nanoscale magnetism.

1. **INTRODUCTION**

Nanomaterials are widely investigated as they own abundant incredible properties distinct with their bulk structures [1-2]. The large surface area and quantum confinement effect of nanomaterials give the main contribution of those distinctions. Since great progress in synthesized technology of two-dimensional (2D) materials provides a new playground for scientist. The heterostructure of 2D materials with different layers or in the same layer have attracted many attentions [3-7]. The 2D quantum dot (QD) is a zero-dimensional structure, the movement of charge carrier in a 2D QD is more limited than a traditional three-dimensional QD and has more strongly quantum confinement effect. It offers an opportunity to explore the quantum confinement effect of QD in different dimensions. 2D QD also carry on the character of traditional QD, such as flexible, easy tuning in electronic and optical properties [8-14]. Moreover, they have easy design by varying different sizes and shapes. The size and shape of a 2D QD determine the electronic and magnetic properties [15-16]. The flexibility of design of QDs provides wide application such as optical devices, electronic devices, biosensing, cancer therapy, and quantum computing [12-24].

Graphene is the first synthesized 2D material and ignite intensely investigation for 2D materials. It possesses many superior properties and behaves as a semimetal, however, the zero gap limit its application. Graphene quantum dot (GQD) is a zero-dimensional (0D) flat structure can be cut from graphene. Because of the strong quantum confinement, GQDs overcome the weak point of graphene and open an energy gap. They own more tunable properties than graphene, such as size, shape, dispersibility, and more active adsorption sites [15, 19-20]. Silicene, the group IV cousin of graphene, has similar honeycomb structure of graphene but the silicon atoms buckled instead of flat structure in graphene. Traditionally, silicon is the primary material used for the electronic devices, new type nanoelectronics made by silicene is expected to reduce the application gap and the industry cost. The investigation on nano-structural silicene is more extensively [25-27]. The silicene transistor had also been achieved in 2015 [28]. Silicene quantum dots (SiQDs) as well as GQDs have great potential for use in a wide range of applications, such as nano optical devices, light emitting diodes, flexible displays, biosensing, bioimaging, therapeutics, drug delivery, photovoltaics, and catalysis [29-32]. Recently, the properties of combined nanomaterials are explored for the effect of heterojunctions, such as the GQDs on one dimensional (1D) $TiO_2$ or 2D $MoS_2$ to enhance the hydrogen evolution reaction [33-34]. GQD on boron nitride nano sheet for microwave absorption [35]. SiQDs in few-layer siloxane for UV−visible absorption and PL emission [31]. Hydrogenation for graphene and silicene is also a widely studied topic, the band gap, electronic and magnetic properties of the one-side, partially, and fully two-side hydrogenated graphene and silicene [36-39]. The embedded GQDs and SiQDs in graphane and silicane had also been explored for the heterojunction of GQDs and graphane (SiQDs and silicane) [15-16].

As quantum confinement has large influence for the 2D QDs, the size of 2D QD plays a key role for band gap [40]. The band gap can be tuned by the size of 2D QD and is reduced as the size of QD is growing large. As well as size, shape of 2D QD is the other important factor for affecting the band gap. For both GQDs and SiQDs, the hexagonal shape with zigzag edges has the largest band gap, and the triangular shape is with a small band gap. The shape of 2D QD also displays different magnetism. The hexagonal, triangular, and parallelogram shapes are referred to nonmagnetic, ferromagnetic, and antiferromagnetic materials for GQDs and SiQDs, respectively. The magnetic moment of triangular GQDs and SiQDs is as a function of size of QDs [15-16]. The embedded GQDs in graphane and the embedded SiQDs in silicane, we denote as GQDs/graphane and SiQDs/silicane, are composite two-dimensional heterostructures (2DH). These combined 2D QDs also can designed as the QDs array. The inter-dot interaction and the strain effect are also the interesting topics for application for quantum computing and flexible displays. This paper presents the study results of both the 2DH systems: GQDs/graphane and SiQDs/silicane, including the inter-dot interaction, strain effect on the electronic and magnetic properties of the 2D QDs array. The physical properties of GQDs and SiQDs affected by strain are also presented as comparison.

## 2. COMPUTATIONAL DETAILS

### 2.1. Structures

Graphene, graphane, silicene, and silicane all have a honeycomb structure, and contains two carbon or silicon atoms in a unit cell, as shown in Figure 1(a). The two atoms occupy at two different sites, named as A and B sites. The carbon atoms of graphene occupied A and B sites are with the same height, i.e., the plane of graphene is flat. However, it is different in graphane, silicene, and silicane, the carbon or silicon atoms situated at the A and B sites are buckled on the opposite sides of the plane. For graphane, the hydrogen atom adsorbed on the carbon atoms at the A and B sites are also on the opposite side of the graphene. For silicane, the hydrogen atom adsorbed on the Si atoms is similar to graphane and the hydrogen adsorption enhance the buckling. The name of QD is denoted as its shape, for example: QDs with hexagonal shape is referred as hexagonal QDs, and the name of QDs with other shapes adhere to the same rule. The hexagonal GQD with twenty-four carbon atoms named as Hex-24 GQD is drawn in Figure 1(b), there are hydrogen atoms terminated at the edge carbon atoms. The Hex-24 SiQD shown in Figure 1(c) follows the same rule as Figure 1(b). Figure 1 (d) and 1 (e) display the GQDs/graphane and SiQDs/silicane structure, respectively. The GQDs (SiQDs) are in the region of C (Si) atoms without hydrogen atoms, and the GQDs (SiQDs) are embedded in graphane (silicane). The coverage of GQD (SiQD) array in graphane (silicane) are also considered for the inter-dot interaction. The notation of $n \times n$ GQDs/graphane (SiQDs/silicane) denotes the GQD (SiQD) array forms a $n \times n$ periodic condition. The smaller the $n$ value, the denser of the GQD (SiQD) in graphane (silicane). Figure 1 (d) and (e) are the $4 \times 4$ hexagonal GQDs/graphane and SiQDs/silicane, which is referred as Hex-$N$ GQD/graphene (SiQD/silicane). $N$ denotes the number of C (Si) atoms inside the GQD (SiQD), and the inter-dot distance ($D$) is shown in Figure 1 (d) and (e). The QDs we considered are all with zigzag edges.

We considered a 2D homogeneous strain and two types of uniaxial strains for the QDs and QDs array systems. The homogeneous strain is an equal biaxial strain, the structure retained its symmetry under such a strain. The homogeneous strain (H-strain $\mathcal{E}_H$) is defined as $\mathcal{E}_H = (a - a_0)/a_0$, where $a$ ($a_0$) is the deformed (undeformed) lattice constant of the QDs array system. The uniaxial strain is defined as $\mathcal{E}_u = (L - L_0)/L_0$, where $L$ and $L_0$ correspond to the deformed and undeformed unit lengths of the applied system in a specific direction. The strain applied in the zigzag direction is named as Z-strain, while the deformation in the armchair direction is called A-strain, as shown in Figure 1 (f).

## 2.2. Calculation Method

The calculations were performed using a first-principles method based on the density functional theory [41] and the generalized gradient approximation [42]. The Vienna *ab initio* simulation package was used. The potentials adopted were based on the projector augmented-wave method [43-44] and were in the Perdew–Burke–Ernzerhof functional form both for the carbon, silicon, and hydrogen ions [45]. The wave functions were expanded by plane waves. Moreover, the energy convergence was tested and the cutoff energy of 500 eV and 312.5 eV was used in the GQDs and SiQDs systems, respectively. Spin polarization was included. Spin-orbit coupling was also considered, but its effect was marginal and could be ignored in this system. The Monkhorst–Pack scheme was adopted for k-point sampling. The k-point mesh was tested up to $15 \times 15 \times 1$, and the k-mesh used depends on the size of supercell of the QDs array systems. One k-point was chosen for the QDs. The system was relaxed until the Hellmann–Feynman force was less than 0.01 eV/Å. The system was simulated using a supercell with a periodic boundary condition in the *x–y* plane and a vacuum layer in the z-direction. The 8 × 8 and 12 × 12 supercells in the x–y plane were considered, which had as many as 256 and 576 atoms, respectively. A vacuum layer thickness of 20 Å was used to prevent the artificial atomic interaction between adjacent supercells.

The strain energy $E_{strain}(\mathcal{E}_u)$ and the inter-dot interaction energy $\Delta E(D)$ are defined as following:

$$E_{strain}(\mathcal{E}_u) = E(\mathcal{E}_u) - E_0, \qquad \text{-------------------(1)}$$

$$E_{dot}(n \times n) = E(n \times n) - (2n^2 - N)E_{Si-H}, \qquad \text{-------------------(2)}$$

$$\Delta E(D) = E_{dot}(n \times n) - E_{dot}(12 \times 12), \qquad \text{-------------------(3)}$$

where $E(\mathcal{E}_u)$ and $E_0$ are the total energy with and without strain. $N$ is the number of carbon or silicon atoms in a QD. $E_{dot}(n \times n)$ and $E(n \times n)$ are the dot energy and the total energy of QD with $n \times n$ periodic structure, respectively. That is the dot energy of a QD with $12 \times 12$ periodic structure is referred as the reference.

## 3. RESULTS AND DISCUSSION

### 3.1. Structural properties.

Firstly, we examine the strain effect for the GQDs/Graphane and SiQDs/Silicane array with various periodic conditions from the point of view of energy. The strain energy ($E_{Strain}$) as defined in equation (1) are shown in Figure 2 (a)-(c). The upper part and lower part in Figure 2 are $E_{Strain}$ of the GQDs/Graphane and the SiQDs/Silicane systems, respectively. Figure 2(a) is the $E_{Strain}$ of hexagonal GQD/Graphane and SiQD/Silicane with six atoms each QD, and five periodic conditions are shown. The energy in an equivalent supercell for different periodic array is used. The H-, A-, and Z-strain are investigated, only the $E_{Strain}$ of H-strain is shown as the behavior of strain energy under strains are similar, but the $E_{Strain}$ under A- and Z-strains is about half of that under H-strain. The $E_{Strain}$ are asymmetric under compression and tension, $E_{Strain}$ under compressive strain is larger than that under tensile strain. This behavior is similar to the that of graphene, graphane, silicene and silicane (Insets in Figure 2(b)). The $2 \times 2$ periodic dot array has largest $E_{Strain}$, the larger the periodic distance is, the smaller the $E_{Strain}$ for both QD array systems. Because fully hydrogenation reduce the hardness of the system, the smaller periodic distance related to dense QD and has larger area without hydrogen adsorbed. Figure 2 (b) presents the $E_{Strain}$ of hexagonal QD with various sizes in $6 \times 6$ periodic structure. Large QD in both systems has larger $E_{Strain}$ as the strain greater than -2%. The difference between the $E_{Strain}$ of the triangular GQD with four and sixteen atoms is tiny, and it is similar for the SiQDs under tensile strains. The $E_{Strain}$ of triangular SiQDs/silicane with larger QDs ($N \geq 16$) is smaller than the Tri-4, as the hydrogenation part in the SiQD/Silicane share more strain energy than that of the part without hydrogen. It is interesting that the actual size of QD is larger than the size that should be under larger compressive strain. It reveals the hydrogenated Si part share some strain on the portion of Si atoms without hydrogen. The strain sharing is more significant for SiQD/Silicane system than GQD/Graphane system. The GQD/Graphane system is harder to be deformed than SiQD/Silicane system.

The interaction energy defined as in equation (3) indicates that the interaction depends on the shape of QD and the species of QD. The inter-dot interaction of SiQD system is different from that of GQD system as plotted in Figure 2(d). There interaction energy between the Hex-6 GQDs in graphane are much smaller than the Tri-4 GQD about its one tenth value. The interaction between the Tri-4 GQDs is strongest in the four cases in Figure 2(d) and shows attractive interaction. When the interdot distance $D$ (as shown Figure 1 d and e) between Hex-6 GQD in graphane is less than 14 Å, the interaction becomes repulsive, and the $D$ decreases the interaction energy increases dramatically as $D$ is less than 5 Å. The $\Delta E$ of Hex-6, Tri-4 SiQDs in silicane is in the region between -0.04 to 0.08 eV. The interaction of SiQDs is firstly attractive then turns to repulsive when $D$ is less than 6.67 and 4.7 Å for the Hex-6 and Tri-4 SiQDs, respectively. The interaction energy between the SiQDs in silicane is smaller than the GQDs in graphane with around one tenth value.

**3.2. Strain effect for magnetism.**

The magnetic properties for QDs are both shape and size dependent, however, the shape dominates the spin density distribution of a QD. The triangle SiQDs and GQDs exhibit as magnetic alternating with spin up and spin down on the adjacent Si or carbon atoms, while the parallelogram SiQDs and GQDs are antiferromagnetic [15-16]. The magnetic property of both 2DH systems can be affected by strains. The strain effect of magnetism is much stronger on SiQDs/Silicane than on GQDs/Graphane. For the four Si atoms inside Tri-4 in silicane, the magnetic moment decreases about 0.1 $\mu_B$ under 10% tensile strain (Figure 3a). Tri-4 in silicane transforms from magnetic semiconductor to nonmagnetic one under compressive strain of 7% except for the $2 \times 2$ periodic condition. Moreover, the Tri-4 with $2 \times 2$ periodic condition begins to lose its magnetism under strain of -5%. It indicates that the strain sharing alters the magnetic properties of a SiQD in silicane. In Figure 3(b), total magnetic moments ($\mu_{tot}$) of the four triangular dots in the $6 \times 6$ periodic condition are presented. The effect on $\mu_{tot}$ under H-strain is stronger than the other two uniaxial strains (A- and Z- strains). Tri-16, and Tri-36 SiQDs in silicane lose their magnetic moment under compressive strain around 5%. Differently, the Tri-25 SiQD firstly convert from magnetic to nonmagnetic semiconductor under H-strain around -3%, and the transformation to nonmagnetic is slower under compressive A- and Z-strains.

**3.3. Strain effect for electronics of GQDs/Graphane.**

The energy gap ($E_g$) of Hex-6 and Tri-4 GQD/Graphane with various periodic conditions are presented in Figure 4, the HOMO-LUMO gap of single Hex-6 and Tri-4 GQDs, and the band gap of graphene are also depicted in Figure 4 as comparison. The $E_g$ of Tri-4 GQDs are smaller than that of Hex-6 GQDs, and less sensitive to strains. The $E_g$ unsensitive with strain of Tri-4 GQD arrays is just like the character of single Tri-4 GQD, but the $E_g$ of Tri-4 GQD arrays show more reactive to strains than that of single Tri-4 GQD. The 3x3 Tri-4 array is the most responsive to strain among the four Tri-4 GQD arrays. The $E_g$ increases as the strain and the variation of $E_g$ from -7% to 10% is about 0.80 eV (0.48 and 0.54 eV) for H-strain (A- and Z- strain), respectively. The increased tendency of $E_g$ is due to the strain effect of graphane. For $2 \times 2$ GQD array, the energy gap of Hex-6 is almost triple value of the Tri-4. However, the $E_g$ of both Hex-6 and Tri-4 in the of 2x2 GQD array is not sensitive to the strains until the strain less than -5%, the $E_g$ of Hex-6 of $2 \times 2$ array dropped around 0.3 eV. The energy gap of single Hex-6 GQD decreases as the H-strain increases. The $E_g$ of Hex-6 GQD arrays increases as the strain increases, that is quite different with the single Hex-6 GQD. Because the HOMO (LUMO) of Hex-6 GQD mix with the valence and conduction bands of graphane, the valence band maximum (VBM) and conduction band minimum (CBM) of the GQDs/Graphane system are dominated by graphane rather than the Hex-6 GQD (Figure 5 a-c). As the conduction band of graphane raising and LUMO of Hex-6 GQD becomes the lowest conduction state under tensile strain, then the increasing rate of $E_g$ is reduced under H-strain. Moreover, the $E_g$ becomes decreasing slowly both under A- and Z-strains, as the LUMO of Hex-6 GQD descendent.

Figure 6 displays the $E_g$ of various shapes and sizes GQDs $6 \times 6$ periodic array and that of the single GQD with hexagonal, triangular and parallelogram shape are also drawn for comparison. As the $E_g$ of QD is shape dependent, Figure 6 shows the hexagonal GQD and GQD 2DH system owns higher $E_g$ than that of the triangular and parallelogram GQDs and GQD arrays. The $E_g$ of GQD array with larger dot size has smaller energy gap. Hex24, Hex-54, Tri-36, and Par-8 single GQD all have larger energy gap than that of GQD array. The $E_g$ of Hex-24 and Hex-54 GQD arrays monotonically deceases as the H-strain increases except of the Hex-54 GQD array under H-strain from -7% to -5%. Differently, the energy gap of hexagonal GQD and GQD arrays declines as the GQD arrays under both compressive and tensile A- and Z-strains. For Hex-24 and Hex-54 arrays, the trend of energy gap under strains is similar to that of single Hex-24 GQD. Because the HOMO and LUMO of Hex24 (Hex54) are located between the CBM and VBM of graphane, the energy is dominated by the GQD part rather than the graphane part in the GQD/Graphane system, as can be seen in Figure 7a. The difference between the largest to smallest $E_g$ of Hex-24 ($\Delta E_g$) are 0.44, 0.56, and 0.55 eV under H-, A-, and Z-strain, respectively, which are smaller than the 1.22, 0.78, and 0.70 eV for single Hex-24 GQD under H-, A-, and Z-strain, respectively. The strain on the GQD in GQD/Graphane system is partly shared by graphane part, as the graphane is more reactive to strain than graphene. The variation trend of energy gap of Tri-36 GQD arrays under strains is different to the single Tri-36 GQD. The $E_g$ of Tri-36 GQD array under H-strain is monotonically increasing and the $\Delta E_g$ around 0.24 eV is greater than that under A- and Z- strains and single Tri-36 GQD under strains. However, the $\Delta E_g$ of Tri-36 under A-and Z-strains are around 0.1 eV and that of Par-72 under strains is less than 0.1 eV. The energy gap of Tri-36 and Par-72 are dominated by GQD because their HOMO and LUMO also is at the band gap region of graphane. However, the graphane part also has small influence on the electronic structures of the 2DH systems. The Tri-36 GQD array is a magnetic semiconductor (Figure 7b), the spin orientation of valence and conduction states of triangular GQD array in the low energy region are opposite. The magnetism of triangular GQD array is still retained under strains from -7% to 10%. The valence states (flat bands in the low energy region) of Tri-36 GQD array seem to be little broadening under large compressive and tensile strains, while conduction states become broadening and are pushed upward as H-strain increases. The difference between the $\Delta E_g$ of Par-72 GQD array and single GQD are smaller than the Tri-36 and Hex-24 2HD systems.

**3.4. Strain effect for electronics of SiQDs/Silicane.**

The energy gap of three shapes of SiQDs/Silicane systems with various periodic conditions under three types of strains are depicted in Figure 8. The single SiQD of the three shapes are also drawn for comparison. The $E_g$ of Hex-6 and Tri-4 SiQD arrays show distinct strain behavior with that of both the single SiQD and silicane under strains. The energy gap of silicane increases monotonically to maximum as H-strain increases to 1% and then decreases linearly as H-strain continuously increases. The $E_g$ of silicane decreases near linearly under both the compressive and tensile strain for A- and Z-strains. The Hex-6 single SiQD exhibits monotonic decline as H-strain increases. For A- and Z-strains, the $E_g$ of single Hex-6 SiQD also decline near linearly both as compressive and tensile strains increase. However, the $E_g$ of Hex-6 SiQD arrays ascends as strain increases, next goes into a slowly growing region, and reaches its maximum, then descends. The maximum $E_g$ of Hex-6 SiQD arrays occurs as H-strain is about 3% and 2% for $2 \times 2$, $4 \times 4$ and $3 \times 3$, $6 \times 6$ SiQD arrays, respectively. Nevertheless, the maximum $E_g$ of Hex-6 SiQD arrays are at zero of A- and Z-strains. The large descendant region occurs in the range of strain less than -3% or greater than 3%. To have further study, the VBM, CBM of silicane and 2DH system, the HOMO and LUMO of single SiQD under strains are plotted in Figure 5b. The VBM of silicane is slightly higher than the HOMO of single Hex-6 SiQD, while the CBM of silicane is overlap with the LUMO of single SiQD and goes downward and exceed the LUMO of SiQD under both compressive and tensile strains. The VBM and CBM of the SiQD 2DH system higher and lower respectively than that of the two individual parts. It indicates that the interaction is much stronger between silicane and SiQD. The band structure of the Hex-6 2DH system with $6 \times 6$ periodic condition under H-strain is drawn in Figure 9a. Surprisingly, the states of SiQD appear in the band gap, the LUMO of Hex-6 SiQD ascends as H-strain increases from -7% to 2%, then mixing in the conduction band. The HOMO of Hex-6 SiQD also moves upward as the tensile H-strain increases. Both the VBM and CBM of silicane part in the Hex-6 2DH system decrease as the strain increases, but HOMO and LUMO of Hex-6 SiQD ascend as the strain increases. Owing to the opposite behavior of band edge of SiQD and silicane parts responding to strain makes the region of higher compressive and tensile strain having lower band gap. The silicane in the SiQD 2DH system shares part of strains of its counterpart without hydrogenation, the strain sharing also contributes to the variation of the $E_g$-strain curve. The variation of size of SiQD is reduced by the strain sharing. The silicane counterpart is over stressed especially in the large strain region and makes the VBM moving upward and CBM moving downward significantly under high compressive and tensile strain, respectively.

As shown in Figure 8 a-c, the HOMO-LUMO gap of single Tri-4 SiQD is unsensitive to strains. However, the $E_g$ of Tri-4 SiQD arrays exhibit the other behavior under strains. The curves of energy gap with strains also have similar tendency to the Hex-6 SiQD array. Nevertheless, the mechanism of Tri-4 and Hex-6 2DH systems is not the same. The energy gap of Tri-4 SiQD arrays increases slowly under the strains in the region of -2% to 5%. The slow varying character is just like the single Tri-4 SiQD under strain, because the HOMO, and LUMO of Tri-4 SiQD is located inside the band gap of silicane, as can be seen in the band structure of Tri-4 SiQD array (Figure 9b). The HOMO and LUMO of Tri-4 SiQD have opposite spin orientation, they are not sensitive to the strain between the range of -3% to 5%. When strain reaches to 5%, the HOMO and LUMO of Tri-4 SiQD mixed with silicane, the CBM of silicane becomes lowered and HOMO of SiQD gets raised. The CBM of silicane is lower than the LUMO of Tri-4 SiQD as the strain is at 10%, furthermore, the CBM gets lower than fermi level and the energy gap becomes zero. For compressive strain part, as strain less then -2%, the VBM is continuously raised up and exceeded the HOMO of SiQD. Moreover, the VBM of silicane breakthrough the fermi level when H-strain reaches to -7%. The Tri-4 SiQD array transfers from a magnetic semiconductor to a nonmagnetic metal as H-, Z-strain and A-strain at -5% and -7% for 2 × 2 respectively (Figure 8 a-c and Figure 3a), but the magnetic transformation is retard until H-strain is at -7% for other periodic condictiones.

The single Par-8 SiQD is antiferromagnetic, however, the Par-8 SiQD array shows nonmagnetic. The antiferromagnetic property of the single Par-8 SiQD can be changed to nonmagnetic as the strain reaches 10 % or less equal to -5%, and -7% for H-strain, and A-, Z-strains, respectively. The energy gap of a single Par-8 SiQD is slow varying in the H-strain range of -3% to 5% and the $E_g$ drops from 1.00 eV to 0.59 eV as the H-strain increases from 5% to 10%, while the variation of $E_g$ is small for the SiQD under A- and Z-strains. As drawn in Figure 9c, the Par-8 SiQD array mix the energy level of SiQD and band of silicane, both the conduction and valence band of silicane part in the SiQD array descends as the H-strain increases. The conduction band of silicane part in array is lowered under compressive A-strain and keep the near same height under compressive Z-strain. The strong mixing of band in silicane part and energy level of SiQD in the SiQD array system leads the energy gap closed at H-strain at 10%.

==========================================================================

The variations of energy gap of three shapes of SiQD arrays under strains are presented in Figure 10, and that of the single SiQD are also drawn for comparison. That includes the cases of hexagonal, triangular shaped SiQD arrays in 6 × 6 periodic condition (Figure 10a to c for hexagonal SiQD and d to f for triangular dot), and the parallelogram shaped SiQD arrays are depicted in 8 × 8 periodic condition. The strain-$E_g$ curves of SiQD arrays all have a slow varying region (SVR), which depends on the dot shapes and size of SiQD arrays, and the strain types. The SVR under H-strain are all smaller than that under A- and Z-strains for hexagonal, triangular and parallelogram SiQD arrays.

For Hex-24 SiQD 6 × 6 array, the SVR is under the H-strain range of -3% ~ 3%, A-strain range of -5% ~ 7%, and -5% ~ 5% for Z-strain. The $E_g$ are spread in the range 0.046, 0.056, and 0.100 eV for H-, A-, and Z-strains, respectively. The SVR of Hex-54 SiQD 6 × 6 array is in H-strain of -5% ~ 5%, A-strain of -7%~5%, and Z-strain of -5%~7%, and for the energy range of 0.051, 0.104, 0.116 eV, respectively. The range of $E_g$ in the SVR increases from approximately 50 meV to 100 meV, and the range of Hex-54 is larger than Hex-24 SiQD array. The variation of $E_g$ of single SiQD is distinct to that of SiQD array. In a SiQD array, the silicane part sheared more strain than the SiQD part, that lead the small variation of energy level of SiQD than of a single SiQD but have larger influence for the band structure of silicane part. The HOMO and LUMO of Hex-24 and Hex-54 SiQD are in the band gap of silicane, the smaller strain sharing of SiQD than silicane lead to less sensitive of HOMO-LUMO gap to the whole strains and from a slow varying region in the strain-$E_g$ curves. Outside the slow varying region, the band structure of silicane strongly mixed with the SiQD, dominates the energy gap of the system and lead to large decay of energy gap as both the compressive and tensile strains get large. Figure 10a displays the band structure of Hex-24 SiQD 6 × 6 array under H-strains. Outside the SVR, the energy level of Hex-24 SiQD is not clear as the strong mixing of energy level of SiQD and band of silicane under -5% H-strain, while the conduction bands of silicane go downward continuously and the energy levels of SiQD move upward, makes the system becomes metallic like as H-strain reaches 10%.

The strain sharing effect on the larger SiQD is more significant than the smaller one. That results the Hex-54 SiQD fast to fall into metallic like material as the H- and A-strain reaches 7%. It also can be observed for Tri-36 becomes as a metal as H-strain reaches 7% but the smaller triangular SiQD array are going metallic like at H-strain of 10%. The SVR of triangular SiQD arrays are more significant than the other hexagonal and parallelogram SiQD arrays. The triangular SiQD array is a ferromagnetic semiconductor changed into a ferromagnet metal under H- tensile strain at 7% (10%) for Tri-36 (Tri-16 and Tri25) and transition to a nonmagnetic metal under H- compressive strain at -3% for Tri25 and -5% for Tri-16 and Tri-36 SiQD array. The band structure of Tri-36 SiQD 6 × 6 array is shown in Figure 10b, it demonstrates the ferromagnetic semiconductor to the nonmagnetic metal transition at -5% H-strain and the transformation of ferromagnetic semiconductor to metal under H-tensile strain larger than 7%. The single parallelogram SiQDs exhibit antiferromagnetic character and the antiferromagnetic does not be changed even under strains for Par-32 and Par-72 SiQDs. The magnetic property of the parallelogram SiQD array is similar to the single SiQD, but it can be changed from antiferromagnetic to nonmagnetic under higher compressive strains. Moreover, the antiferromagnetic semiconductors of parallelogram SiQD array also have transition from a antiferromagnetic semiconductor to antiferromagnetic metal under a 10% H-strain. The former is due to the interaction between SiQD under compressive strains, the latter is because of the sharing strain by the silicane part and the mixing of energy level of SiQD and band of silicane which can be seen in Figure 10c.

## 4. CONCLUSION

We had performed a serious DFT calculations for the deformation effect on the GQD/Graphane and SiQD/Silicane array systems. The electronic structure is mainly composed by the GQD (SiQD) array and Graphane (Silicane), additionally the interdot interaction and the strain sharing effect for graphane or silicane enriched the factors for the manipulation of electronic property. The interdot interaction is stronger for GQD/Graphane than the SiQD/Silicane system, on the contrary, the strain sharing effect is much significant for SiQD/Silicane than GQD/Graphane system. The strain sharing effect push down the conduction band of silicane caused the transition from semiconductor to metal under higher tensile strains. The strain sharing accompanied with interdot interaction results the magnetic to nonmagnetic transition for triangular and parallelogram SiQD array, and the semiconductor to metal transition for hexagonal and triangular SiQD array. The magnetic properties of GQD/Graphane do not change by deformations. The strain effect can used to turning the magnetic and nonmagnetic for the triangular and parallelogram SiQD/Silicane. It can be applied in nanoscale registrations of digital information. SiQDs/Silicane can also be used to design silicon-based microelectronic devices.

FIGURES:

Figure 1. (a) The left panel is top view and side view of graphene and silicene. The right panel is the hydrogenated cases: graphane and silicane. Top view and side view of (b) hex-24 GQD and (c) hex-24 SiQD, the edges are saturated by hydrogen atom. Top view of (d) GQD array and (d) SiQD array, $D$ is the inter-dot distance. (f) The system is under H- (left), A- (middle), and Z- (right) strains.

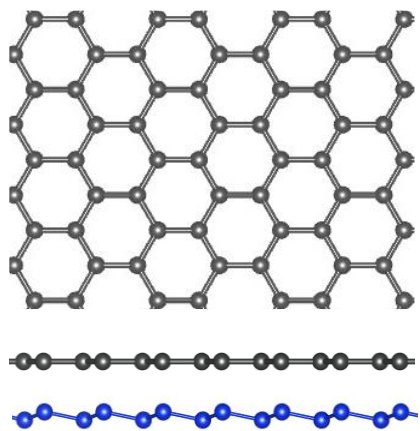
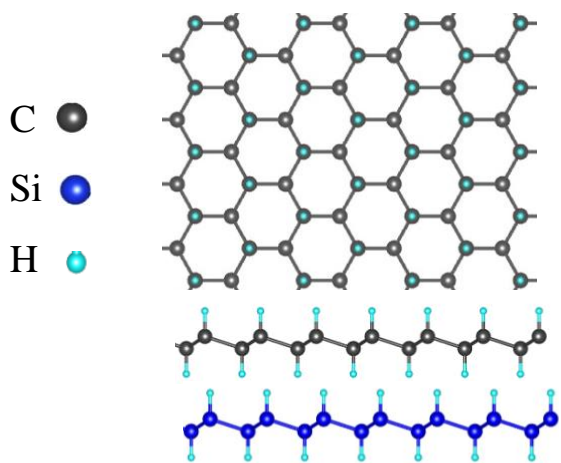
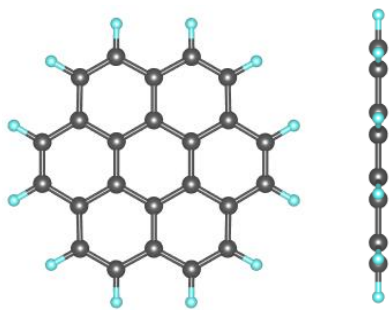
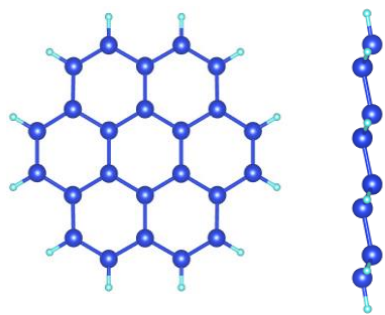
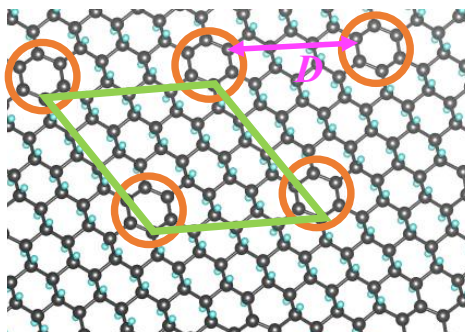
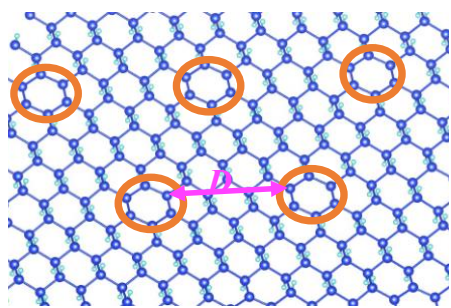
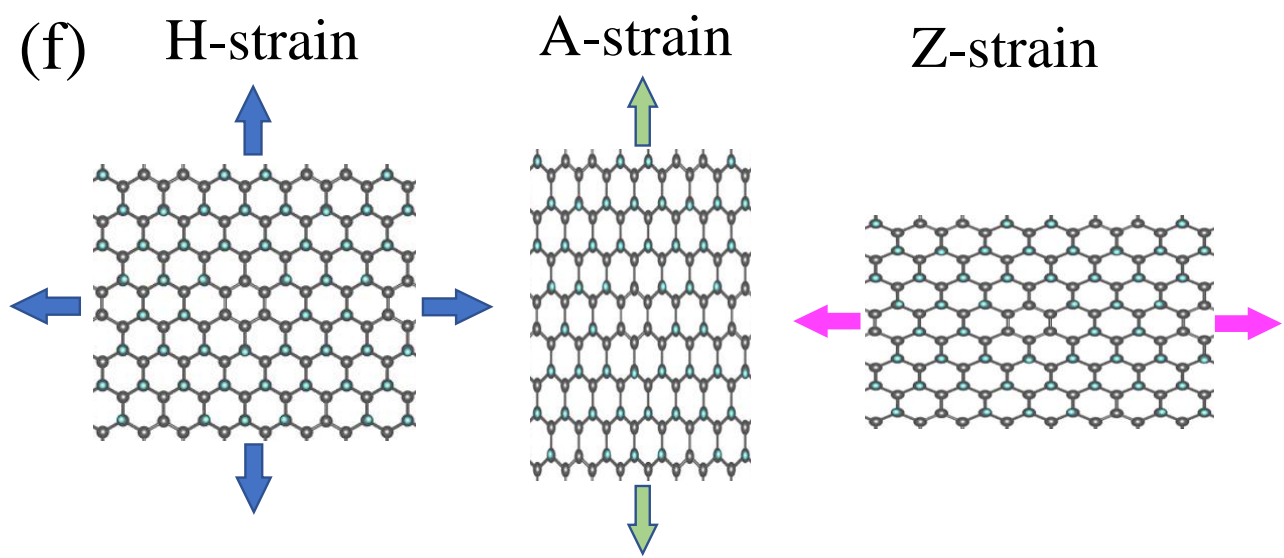

Figure 2. Strain energies with various H-strains (a)-(c), the upper part and lower part are for GQD/Graphane and SiQD/Silicane, respectively. (a) Hex-6 QD array with various array size. (b) The hexagonal dots with $N = 6, 24,$ and $54$ in $6 \times 6$ QD array, the insets shown the strain energy of graphane and graphene (silicane and silicene). (c) Triangular dots with $N = 4, 16$ for GQDs and $N = 4, 16, 25, 36$ for SiQDs in 6x6 array. (d) The inter-dot interaction energy of Hex-6 and Tri-4 QD arrays, the reference is set to $12 \times 12$ array.

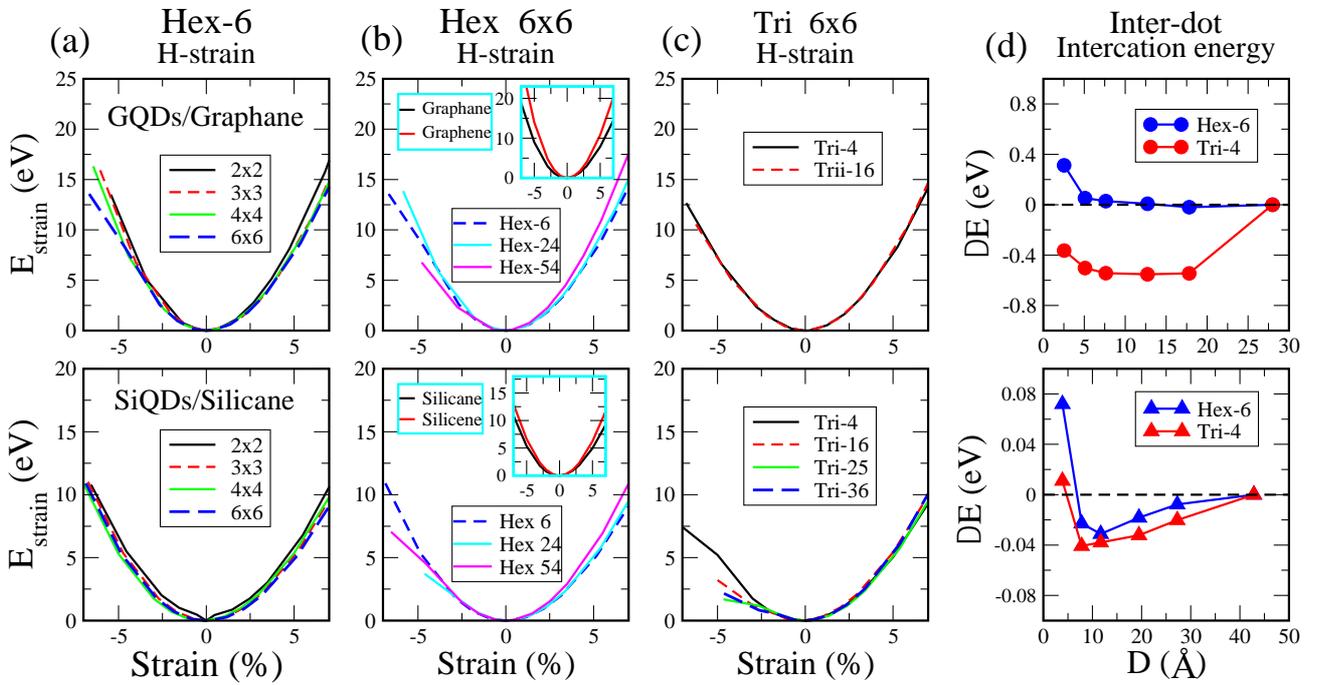

Figure 3. (a) Total magnetic moments of Tri-4 SiQDs in various periodic arrays under strains. (b) Total magnetic moments of triangular SiQDs with $N = 4, 16, 25, 36$ in $6 \times 6$ array under strains. Unit is in bohor magneton $\mu_B$.

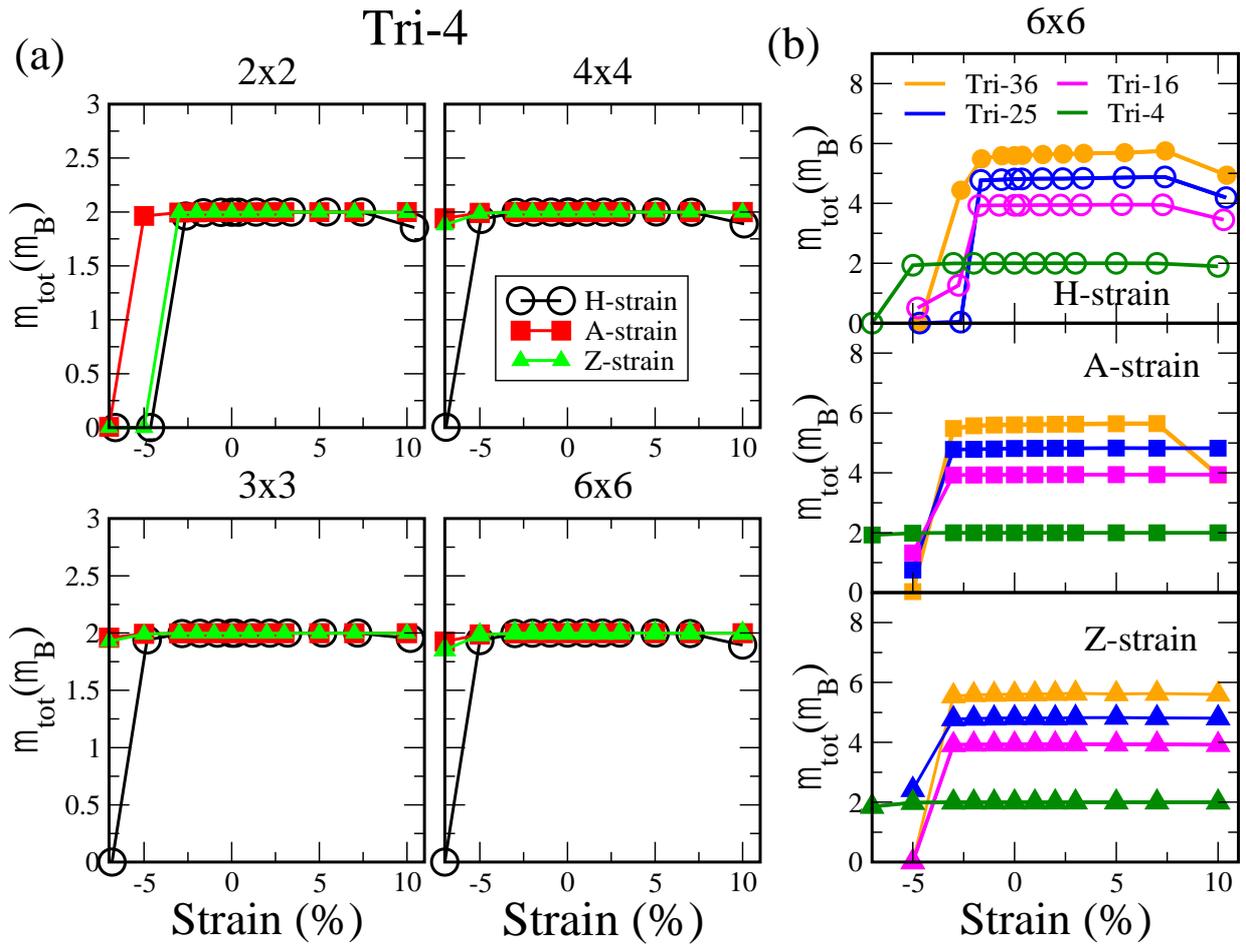

Figure 4. Variation of energy gap of Hex-6 and Tri4 GQD/Graphane 2DH system with four periodic conditions under (a) H-, (b) A-, and (c) Z-strain. Circles and triangles are the energy gap of Hex-6 and Tri4 GQD/Graphane system, respectively. The solid circles and solid triangles are energy gap of single Hex-6 and Tri-4 GQD, respectively. Magenta dashed line are the energy gap of graphane.

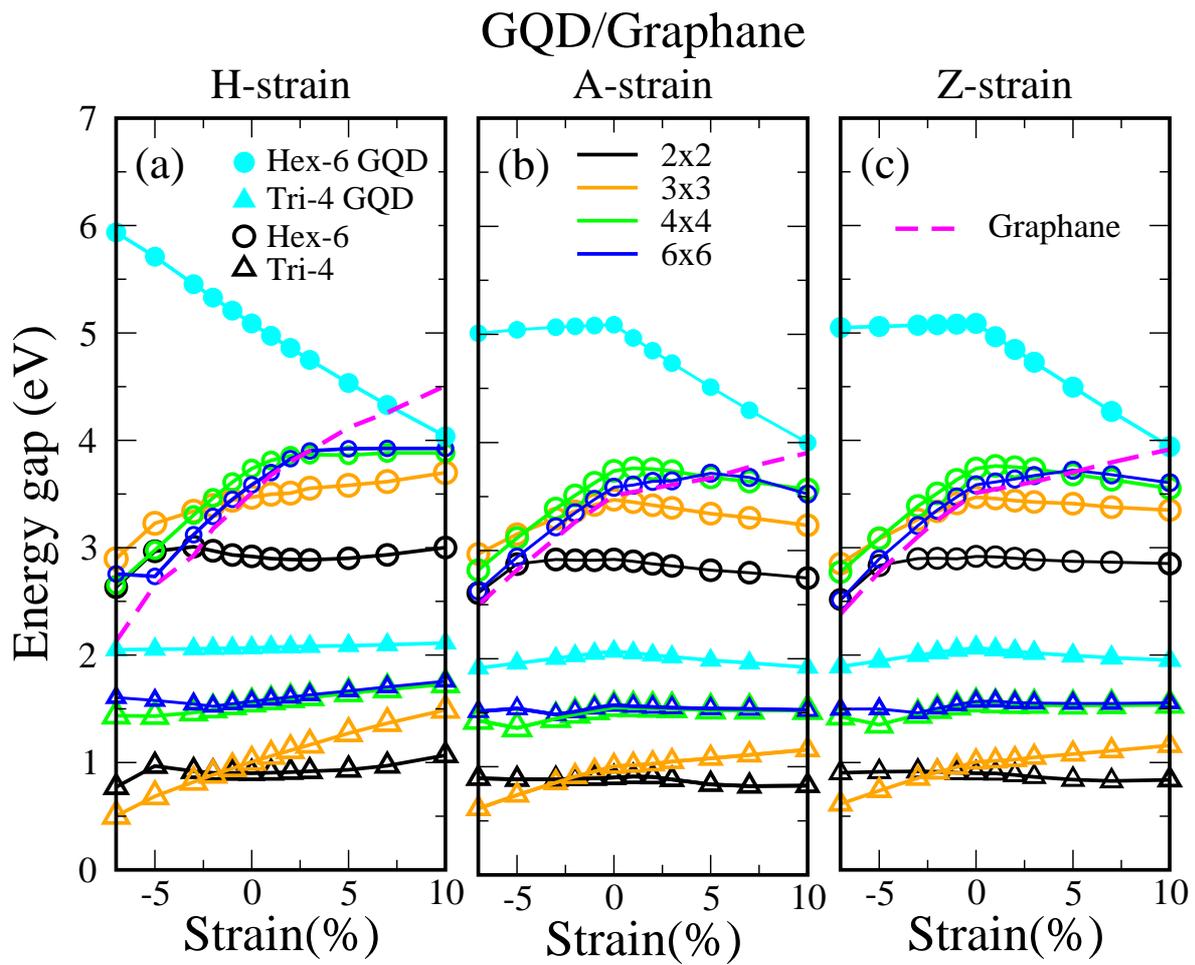

Figure 5 Conduction and valence band edges of 6 × 6 Hex-6 arrays under strains. (a)-(c) is for the GQD/Graphane under H-, A- and Z-strains. (d)-(f) is for the SiQD/Silicane under H-, A-, and Z-strains. The black (blue) circles are the CBM and VBM of Hex-6 GQD (SiQD) arrays, the red (black) dashed lines are the CBM and VBM of graphane (silicane), and the cyan (magenta) solid circles are the LUMO and HOMO of Hex-6 GQD (SiQD). The energy of CBM and LUMO are positive, while the energy of VBM and HOMO is negative.

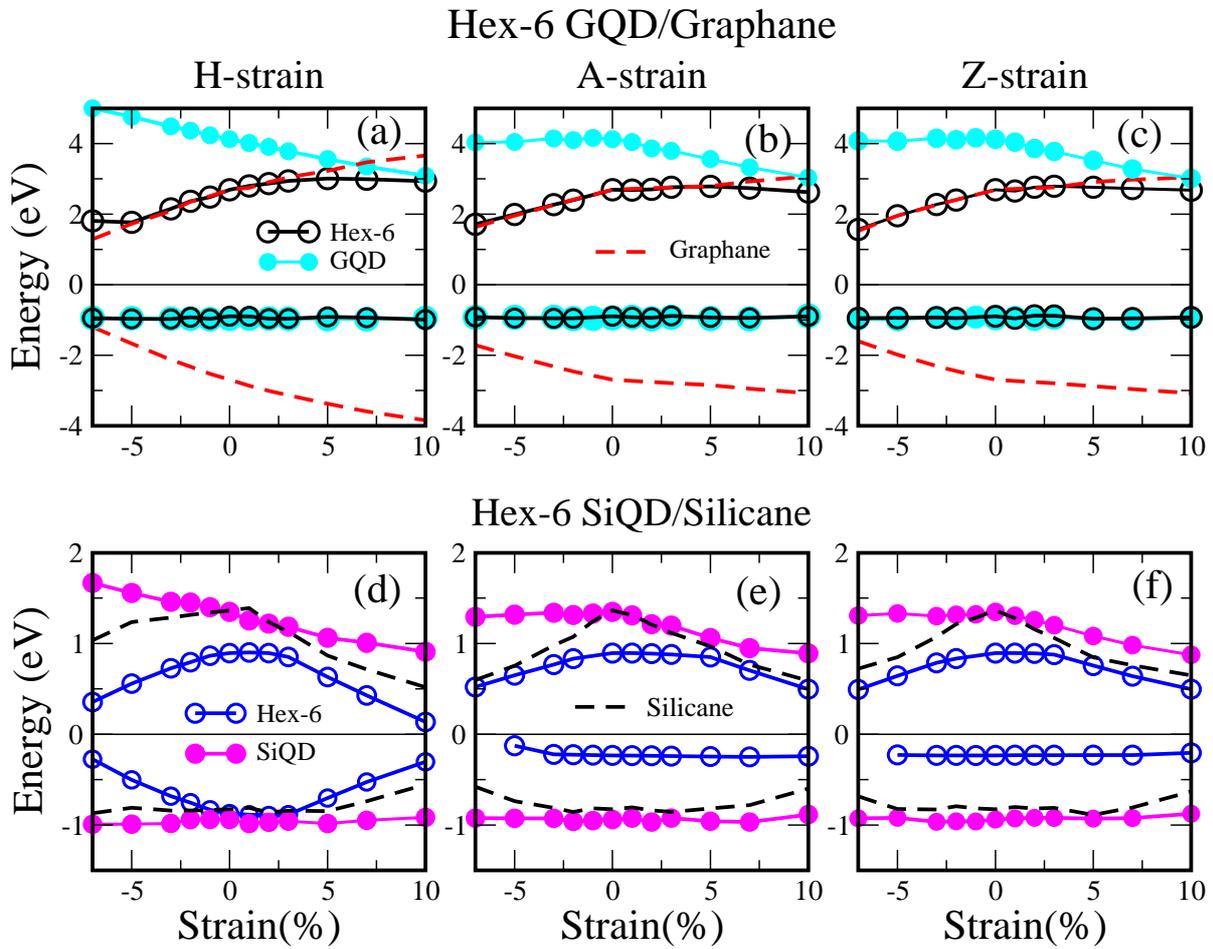

Figure 6. Energy gap of GQD/Graphene with 6 × 6 periodic condition under (a) H-, (b) A-, and (c) Z-strains. The black solid circles, solid triangles, and solid diamonds denote the energy gap of a single Hex-24, Tri-36, and Par-72 GQD, respectively.

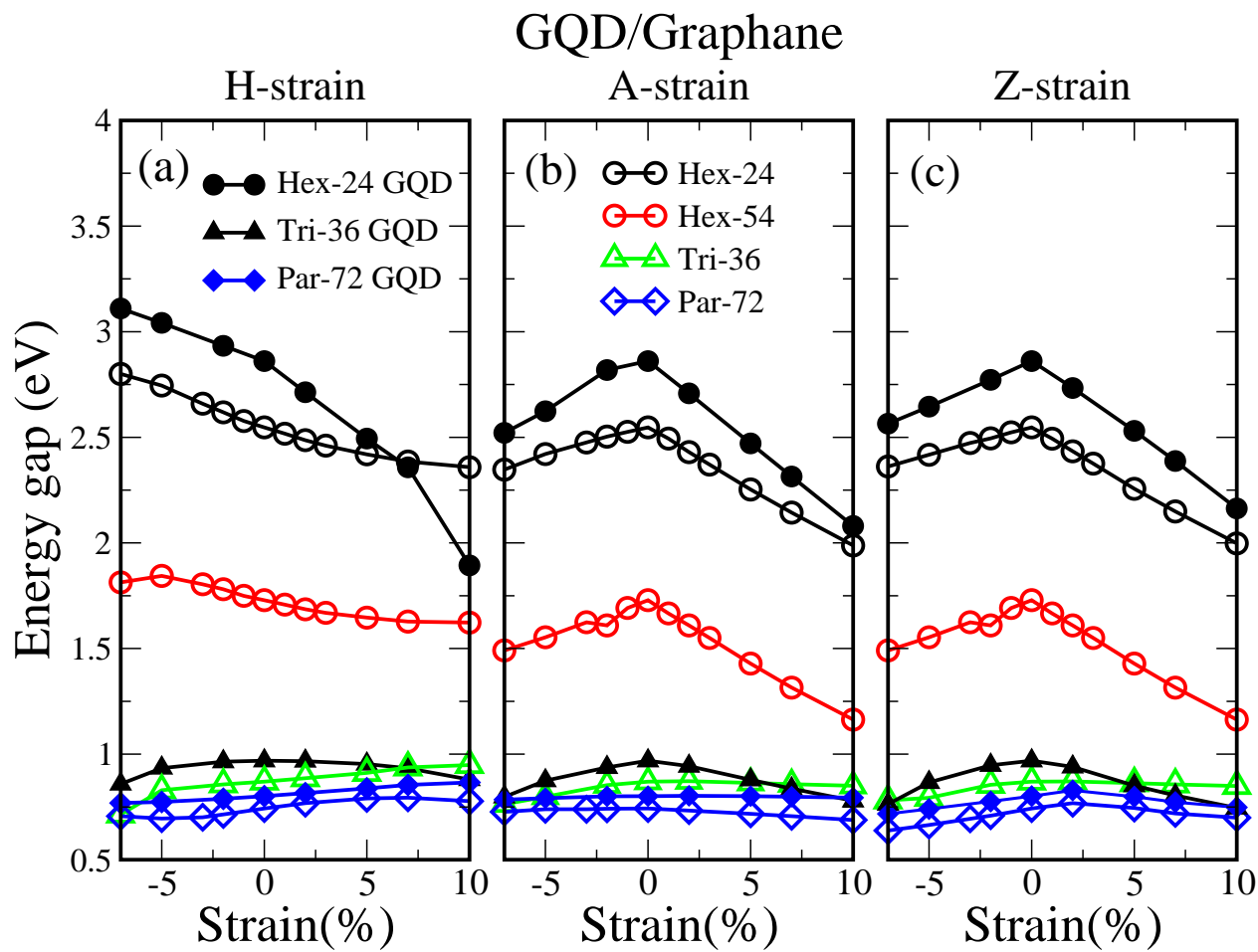

Figure 7. Band structures of (a) Hex-24 GQD 6x6 and (b) Tri-36 GQD 8x8 array 2DH system with various H-strains.

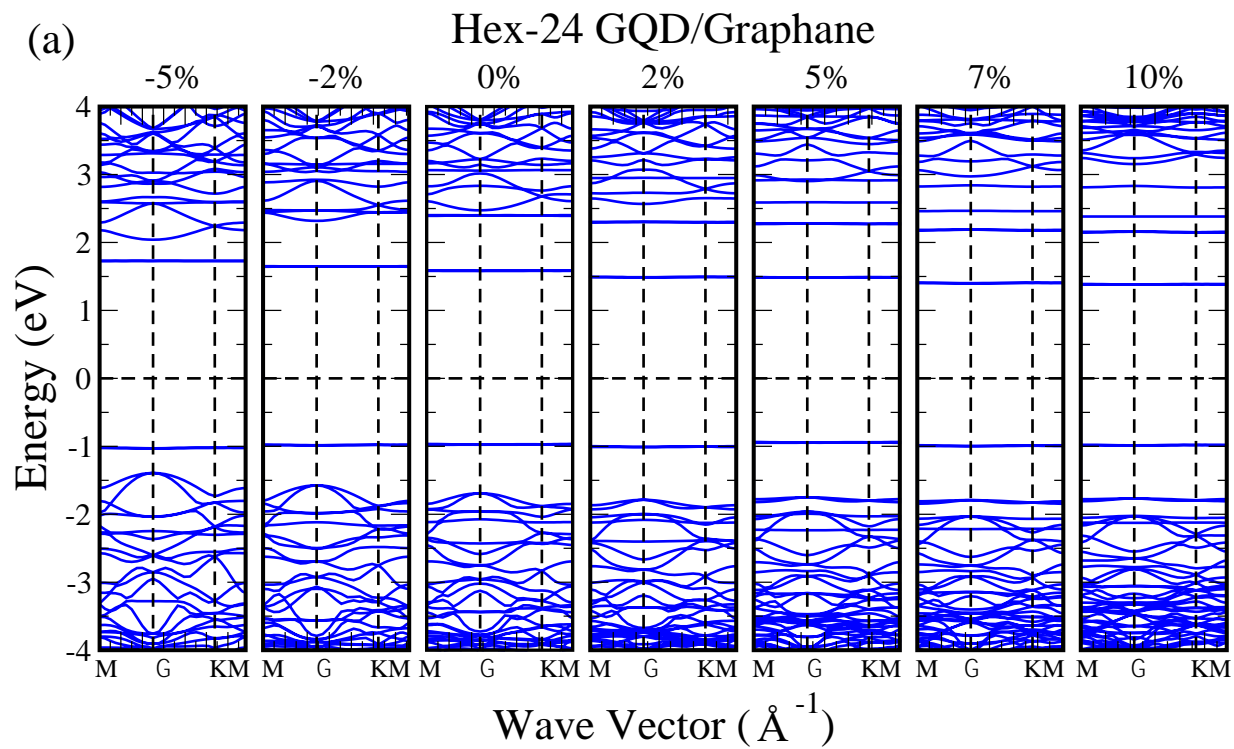

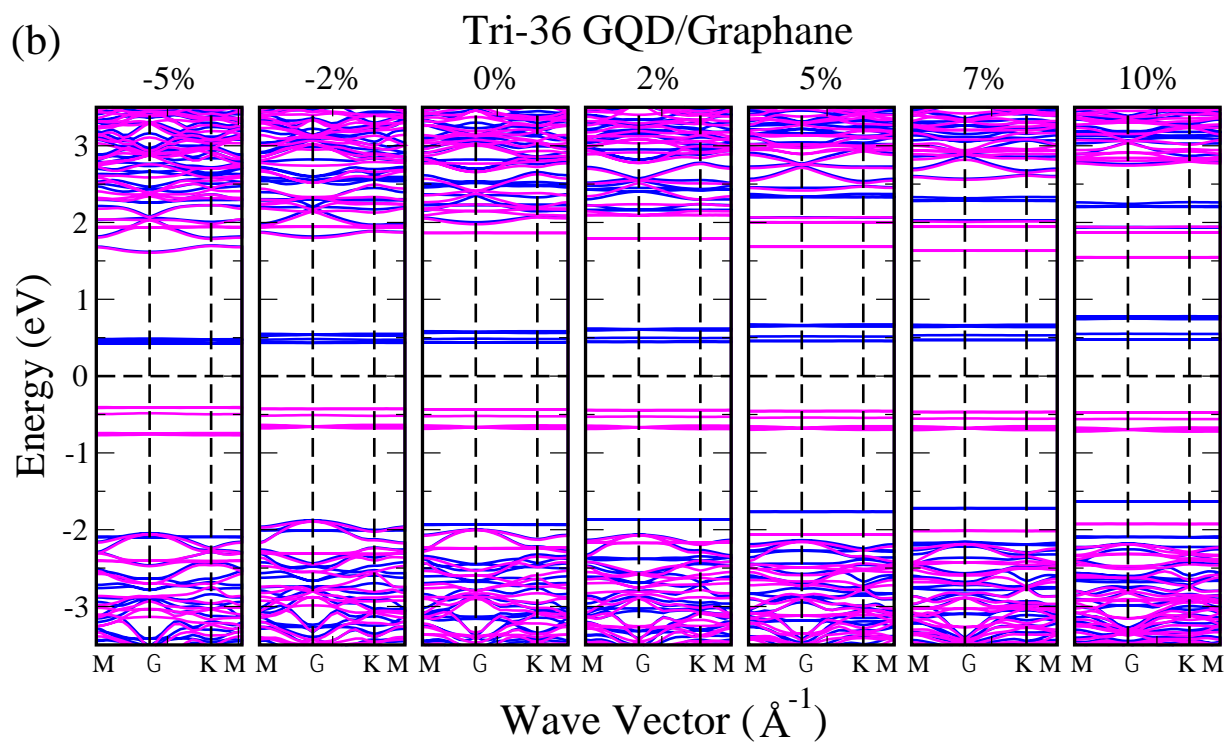

Figure 8. Variation of energy gap of Hex-6 SiQD/Silicane 2DH system with six and five periodic conditions under (a) H- , (b) A-, and (c) Z-strain. (e-f) The energy gap-strain curves of Tri-4, and Par-8 SiQD 2DH systems. Circles, triangles, and diamonds are the energy gap of Hex-6, Tri-4, and Par-8 SiQD/Silicane system, respectively. The magenta solid circles, solid triangles, and solid diamonds are energy gap of single Hex-6, Tri-4, and Par-8 SiQD, respectively. Black dashed lines are the energy gap of graphane under strains.

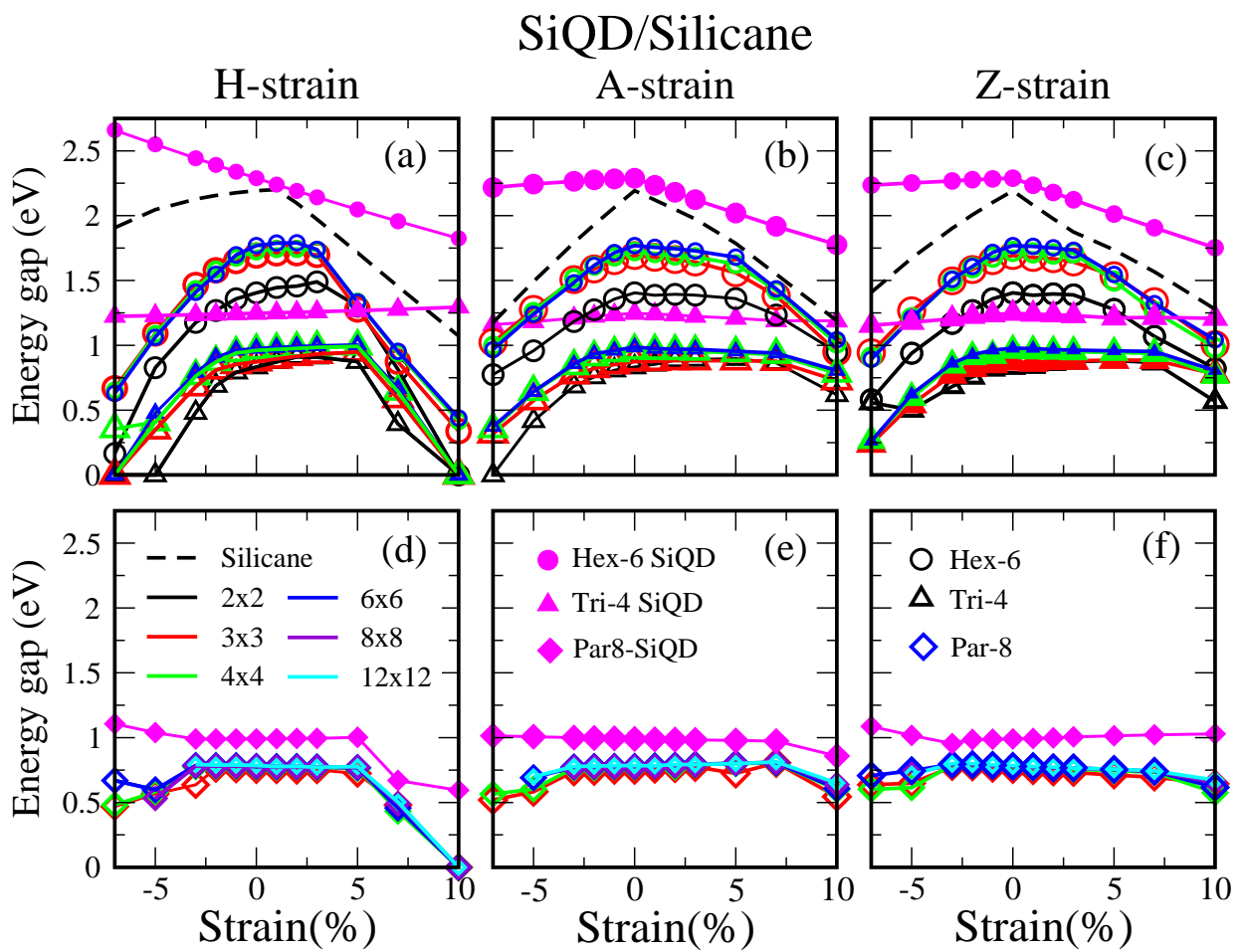

Figure 9. Band structure of (a) Hex-6, (b) Tri-4, (c) Par-8 SiQD/Silicane 6 × 6 array under strains.

AUTHOR INFORMATION

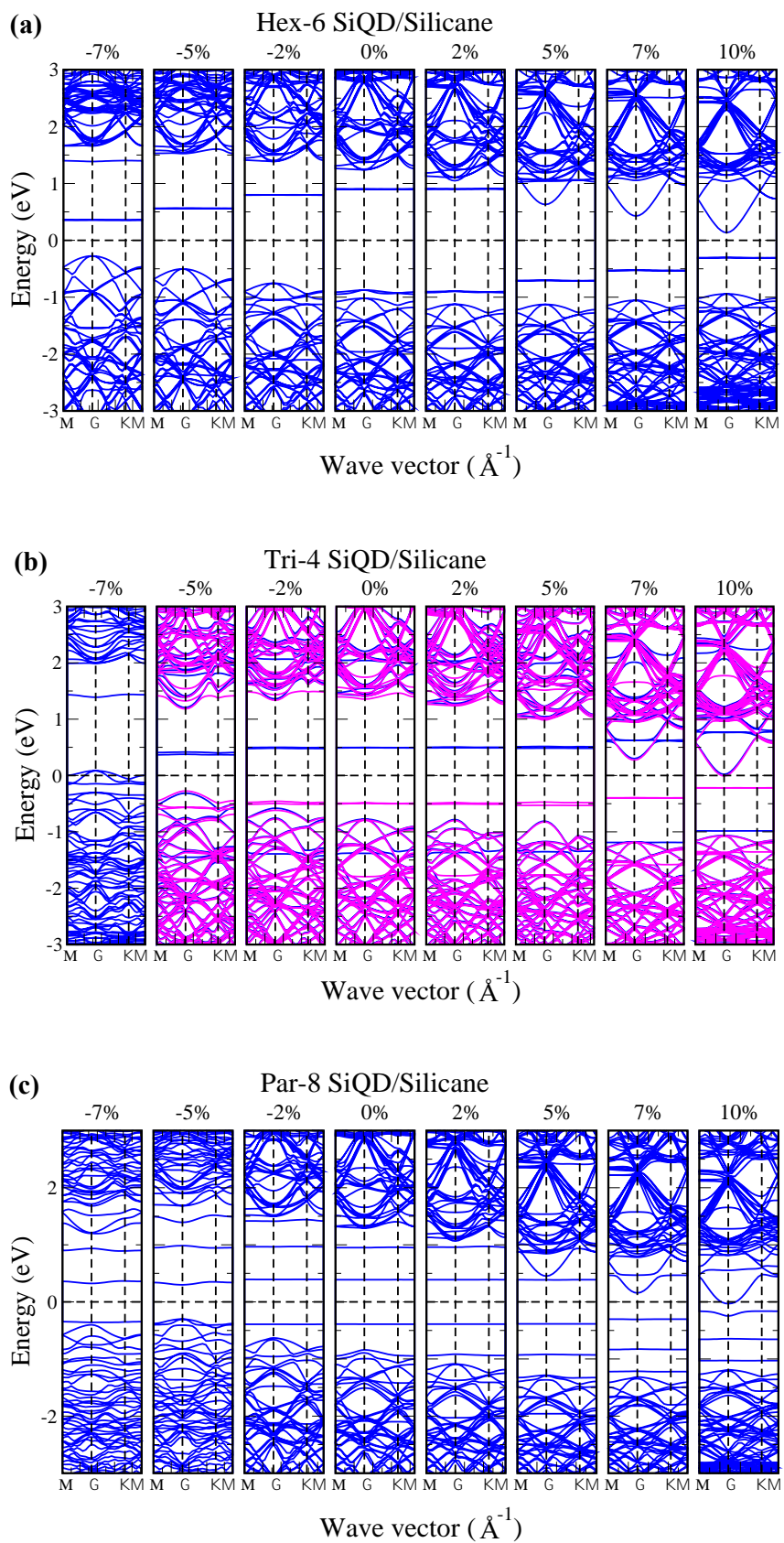

Figure 10. Energy gap of SiQD/Silicane with 6 × 6 periodic condition under (a) H-, (b) A-, and (c) Z-strains. The black solid circles, blue solid triangles, and solid diamonds denote the energy gap of a single Hex-24, Tri-36, and Par-32, Par-72 GQD, respectively.

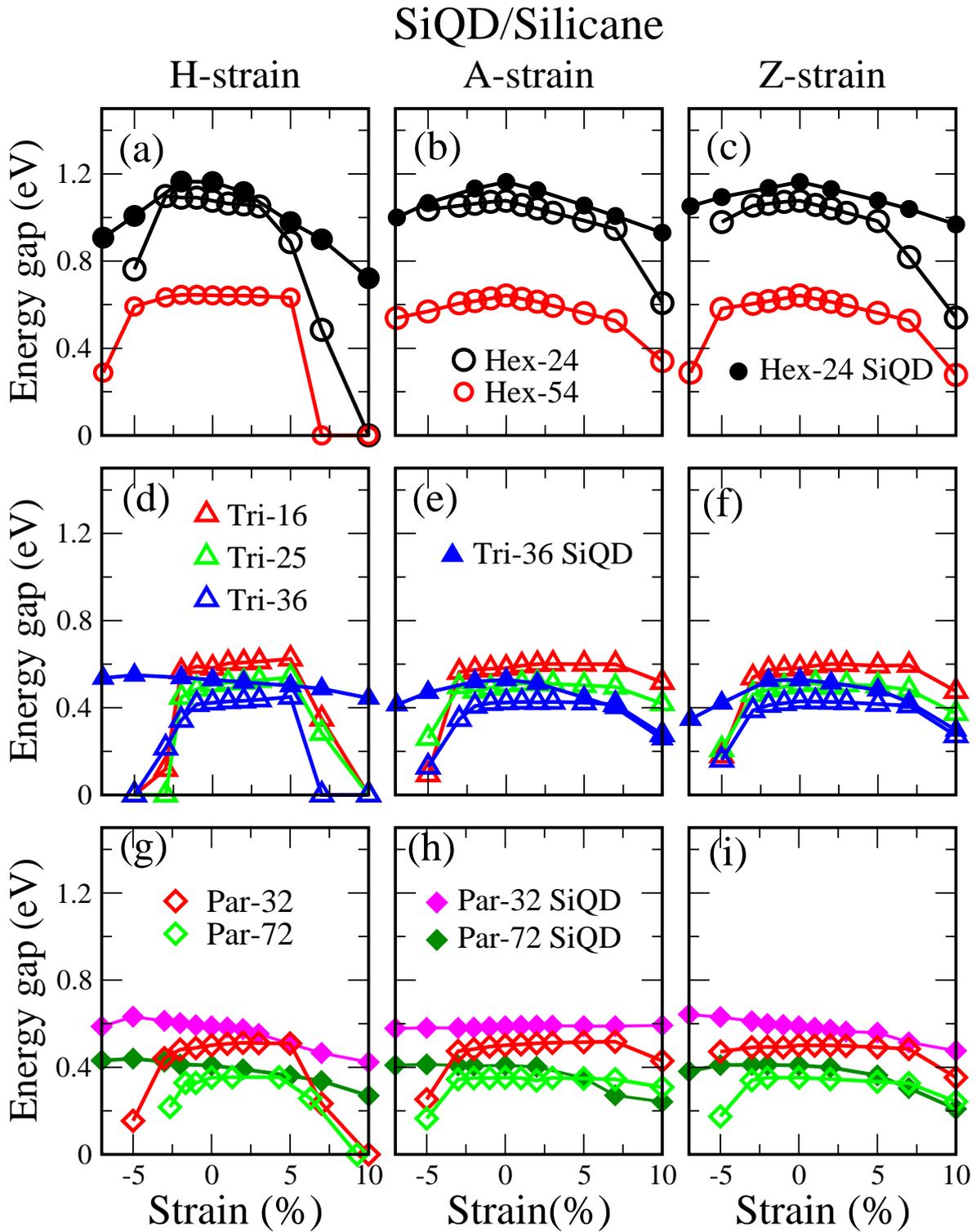

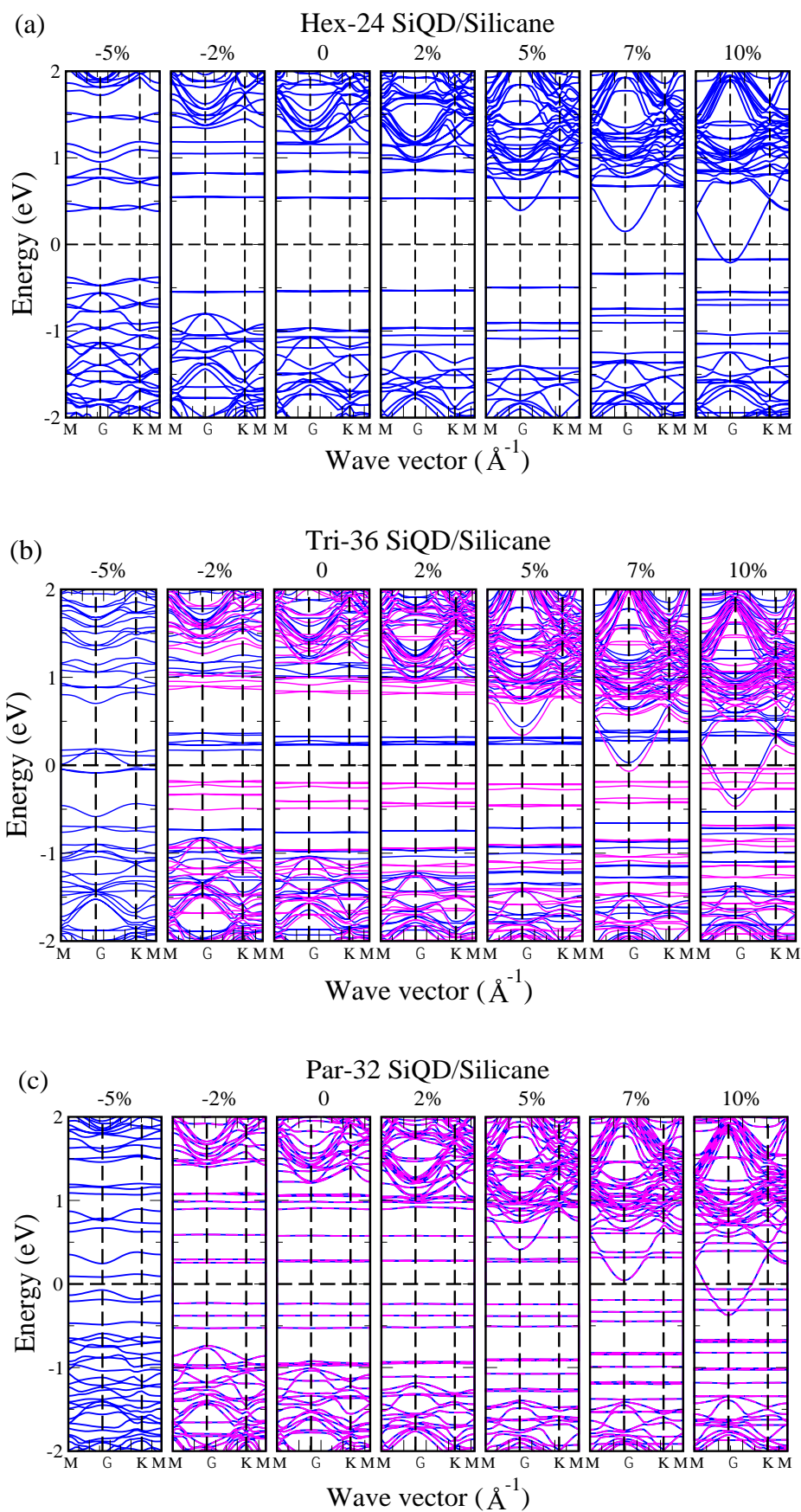

Figure 11. Band structure of (a) Hex-24, (b) Tri-36 (c) Par-32 SiQD/Silicane 6 × 6 array under H-strain.

**Corresponding Author**

*E-mail: brwu@mail.cgu.edu.tw (B.-R. W.).

**Author Contributions**

The work was performed by, and manuscript was written through B.-R. W.

**Funding Sources**

This work was supported by the Cheng Gung University under grant numbers UMRPD5M001.

**Notes**

The authors declare no competing financial interest.

**ACKNOWLEDGMENT**

This work was supported by the Cheng Gung University under grant numbers under grant numbers UMRPD5M001.